\documentclass[twocolumn,amsmath,amssymb,a4paper,superscriptaddress]{revtex4}

\usepackage{graphicx}% Include figure files

\begin{document}

\hyphenation{nano-disc}

%---------------------------

\title{Nanoantenna-enhanced ultrafast nonlinear spectroscopy of a single gold
nanoparticle}

%% Autorenliste und Institute
\author{Thorsten Schumacher}
\author{Kai Kratzer}
\author{David Molnar}
\author{Mario Hentschel}

\affiliation
{Max Planck Institute for Solid State Research, Heisenbergstrasse 1, D-70569
Stuttgart, Germany}
\affiliation{4$^\mathit{th}$ Physics Institute and Research Center SCoPE,
University of Stuttgart, Pfaffenwaldring 57, D-70550 Stuttgart, Germany}

\author{Harald Giessen}
\affiliation{4$^\mathit{th}$ Physics Institute and Research Center SCoPE,
University of Stuttgart, Pfaffenwaldring 57, D-70550 Stuttgart, Germany}

\author{Markus Lippitz}
\email{m.lippitz@physik.uni-stuttgart.de}
\affiliation
{Max Planck Institute for Solid State Research, Heisenbergstrasse 1, D-70569
Stuttgart, Germany}
\affiliation{4$^\mathit{th}$ Physics Institute and Research Center SCoPE,
University of Stuttgart, Pfaffenwaldring 57, D-70550 Stuttgart, Germany}

%%ende Autorenliste

\date{\today}

\maketitle

\textbf{%
Optical nanoantennas are a novel tool to investigate previously unattainable
dimensions in the nanocosmos. Just like their radio-frequency equivalents, 
nanoantennas enhance the light-matter interaction in their feed gap \cite{farahani05jul1,kinkhabwala09nov}. 
Antenna enhancement of small signals promises to open a new regime in  linear and nonlinear
spectroscopy on the nanoscale. Without antennas especially the nonlinear
spectroscopy of single nanoobjects is very demanding \cite{stievater01sep24,hartschuh03sep5,lippitz05apr,langbein05jul1,%
vandijk05dec31,butet10may,hwang09jul2,brinks10jun17}.
Here, we present for the first time antenna-enhanced ultrafast nonlinear optical spectroscopy. In particular, we
utilize the antenna to determine the nonlinear transient absorption signal of a
single gold nanoparticle caused by mechanical breathing
oscillations \cite{vandijk05dec31}. We increase the signal amplitude by an order of magnitude which is in good
agreement with our analytical and numerical models. Our method will find
applications in  linear and nonlinear spectroscopy of nanoobjects, ranging
from single protein binding events \cite{mayer10jun25} via  nonlinear tensor
elements  \cite{maeda05feb4} to the limits of continuum mechanics  \cite{juve10may}.
}

\begin{figure}[b!]
\includegraphics[scale=1]{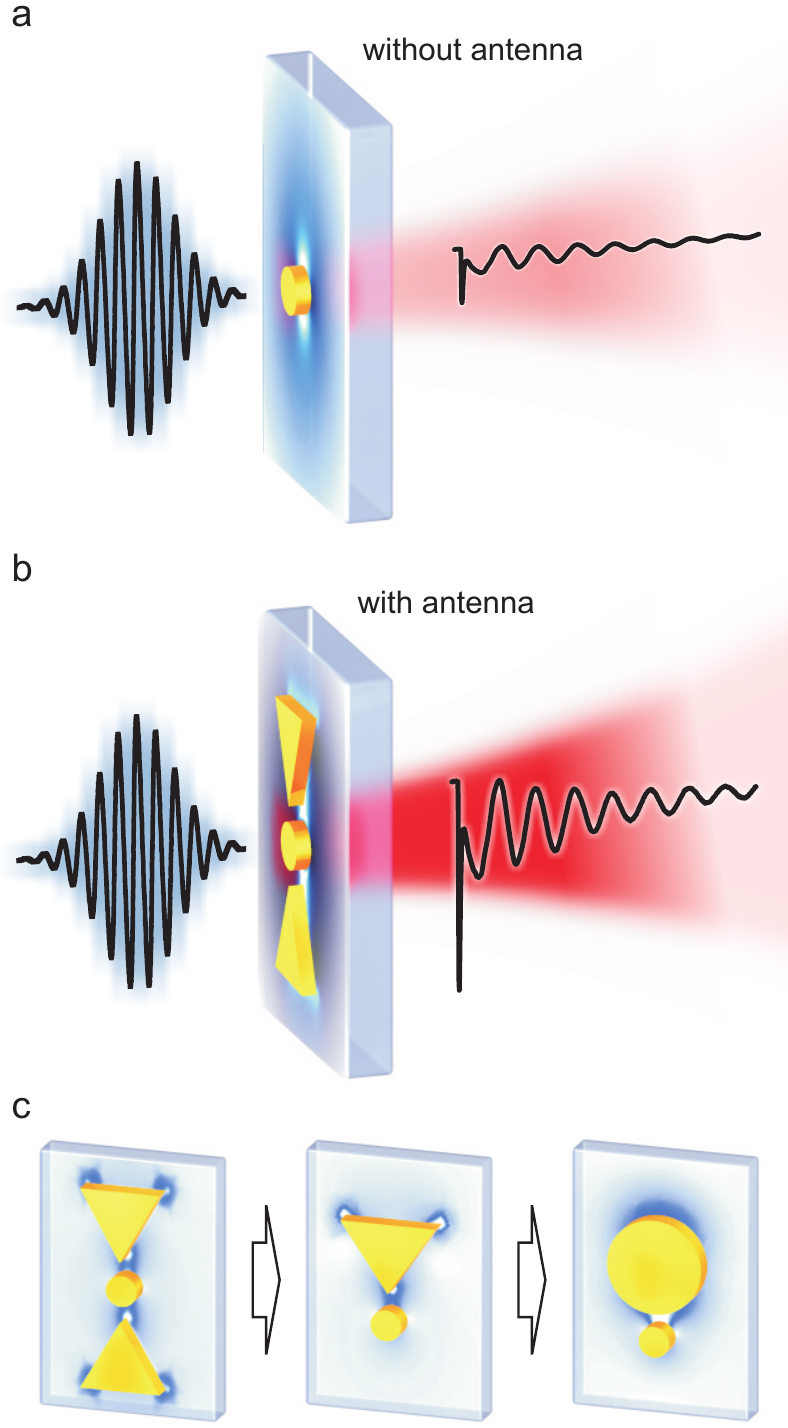}
\caption{\textbf{Illustration of signal enhancement using a resonant optical
nanoantenna:} \textbf{a}, A single nanoobject of few nanometers in size is
probed by an ultrashort laser pulse. The resulting nonlinear response is
extremely weak but carries informations about the physical processes.
\textbf{b}, An optical nanoantenna (represented by the bowtie structure)
enhances the nonlinear response of the nanoparticle by increasing the
light-matter interaction at the nanofocus. \textbf{c}, We demonstrate this effect using a polarization-insensitive dipole antenna, i.e., a single disc. }
\label{fig:intro}
\end{figure}

Nanoobjects with a size between 1 and 100~nm show fascinating properties which
deviate strongly from bulk behavior. The plasmon resonance of metal
nanoparticles and the electron confinement in quantum dots are prominent
examples. However, even with the best preparation methods, the individual
objects differ from each other in size, shape, or local environment, rendering
necessary single object experiments \cite{moerner99mar12}. As the light-matter
interaction strength scales with the number of  electrons involved, the signals
of an individual nanoobject become very weak. Especially the nonlinear signals,
which are already weak for bulk material, become difficult if not impossible to
detect. A resonant optical nanoantenna \cite{footnoteresonant}
that concentrates the optical field on the
individual nanoobject promises enhancement of  such weak nonlinear signals (see
Fig.{~\ref{fig:intro}). In previous work, nonlinear effects were  used to
characterize the antenna itself
\cite{muhlschlegel05jun10,leitenstorfer09dec}. 
\textit{Here we utilize the resonant antenna for spectroscopy and demonstrate nanoantenna enhancement
of the ultrafast transient transmission of a single metal nanodisc}
\cite{vandijk05dec31}. 
This transient transmission spectroscopy is a non-perturbative optical method to investigate mechanical properties at the nanoscale. It is particular suited to test our concept.
A pump-pulse triggers acoustical vibrations. The oscillations lead to
a periodic variation of the particle size and of the plasma frequency via the electron density.
This modifies in turn  the dielectric function \cite{perner00jul24,hodak99nov8}.   In this way, we determine the oscillation eigenfrequency of a tiny
nanodisc and in this way Young's modulus at the nanoscale \cite{hartland06annur}.

\begin{figure}[t]
\includegraphics{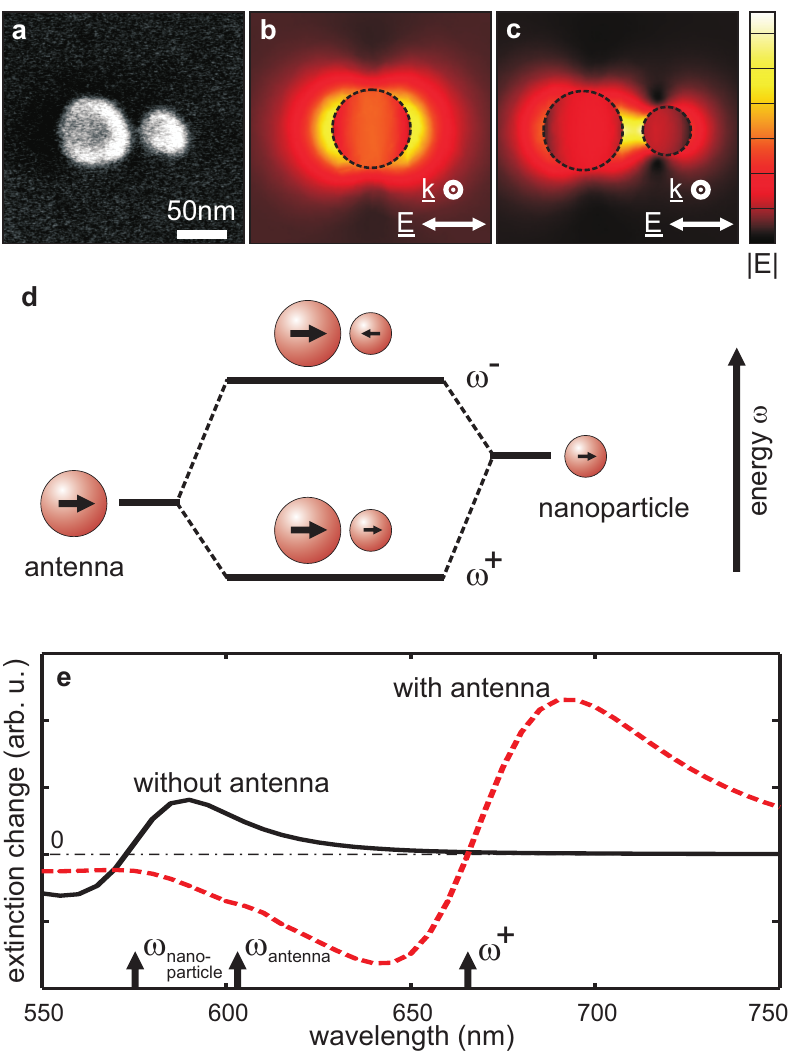}
\caption{\textbf{Basic concept of an antenna for a plasmonic particle:}
\textbf{a}, SEM picture of the structure: nanoantenna (disc diameter 70~nm)
separated from the nanoparticle under investigation (disc diameter 40~nm) by a gap of 15~nm. Both
elements are made by electron beam lithography out of gold (height 30~nm) on a
glass substrate. \textbf{b,c}, Absolute value of the electric field in the top plane of the structure. The wavelength is
in each case adjusted to the resonance. \textbf{d}, A coupled
antenna-nanoobject system has new eigenmodes, determined by plasmon hybridization.
\textbf{e}, The change in extinction  of a coupled system (dashed line) by varying
the nanoparticle's plasma frequency is
enhanced and red shifted, compared with the case of an uncoupled nanoparticle
(solid line).
} 
\label{fig:idea}
\end{figure}

Currently, different kinds of antenna structures are under development in the
optical regime, ranging  from simple dipoles to complex Yagi-Uda
antennas \cite{bharadwaj09xxxxx,muhlschlegel05jun10,merlein08apr,kosako10may,curto10aug20}. In our experiment, we use a
polarization-insensitive dipole antenna, i.e., a plasmonic nanodisc of 70~nm
diameter (Fig.{~\ref{fig:idea}a}). As the nanoobject under investigation we pick a nanodisc of smaller diameter
(40~nm) whose linear optical scattering signals are already very weak and
whose transient absorption is nearly impossible to detect.

The scattering characteristic of both discs is dominated by a dipolar plasmon
resonance. Close to the resonance, the optical
near-field around the disc shows a dipolar pattern with two regions of high
field intensity (Fig.{~\ref{fig:idea}b}). These hot spots act similar to a nano-lens
by concentrating the field in a small volume \cite{li03nov28}. A variation of the dielectric
function in this volume will have a  large
influence on the transmitted field. This is the antenna effect we exploit for 
signal enhancement. In our case, the perturbation of the antenna by the
nanoobject is not negligible. Both antenna and nanoobject together form a
coupled system with common
eigenmodes and an altered field distribution (Fig.{~\ref{fig:idea}c}).

The common eigenmodes of the coupled system can be described in the plasmon
hybridization picture \cite{prodan03oct17,liu10angew} by   dipole-dipole interactions. These coupled modes are the symmetric (\(\omega^{+}\)) and
antisymmetric (\(\omega^{-}\)) combination of the original modes (Fig.{~\ref{fig:idea}d}). If antenna and nanoobject were equal in
size, the symmetric mode \(\omega^{+}\) would carry all the oscillator strength.
In our case, both modes are optically active. However, the high energy
antisymmetric mode \(\omega^{-}\) is shifted into the d-band absorption of gold
and thus damped out. For a polarization direction perpendicular to the symmetry
axis, the dipole coupling and thus the mode splitting is reduced. The strong, symmetric mode becomes the
blue-shifted mode and is damped due to absorption.

In the uncoupled case, a pump-induced variation of the  plasma frequency
of the nanoparticle shifts the
extinction spectrum of the nanoparticle only, resulting in a dispersive feature
in the differential spectrum at the plasmon resonance frequency \(\omega_{\text{nanoparticle}}\) (solid line in
Fig.{~\ref{fig:idea}e}). When antenna and nanoparticle form a coupled system,
the influence of the dielectric constant of the particle
is amplified by the larger oscillator strength of  the coupled system. This results in a larger
variation of the extinction (dashed line in Fig.{~\ref{fig:idea}e}) 
now located at the  coupled antenna mode  \(\omega^{+}\).
Thus a successful
operation of the antenna is characterized by a red shift and an amplification of
the nonlinear signal of the particle.

\begin{figure*}[t]
\includegraphics{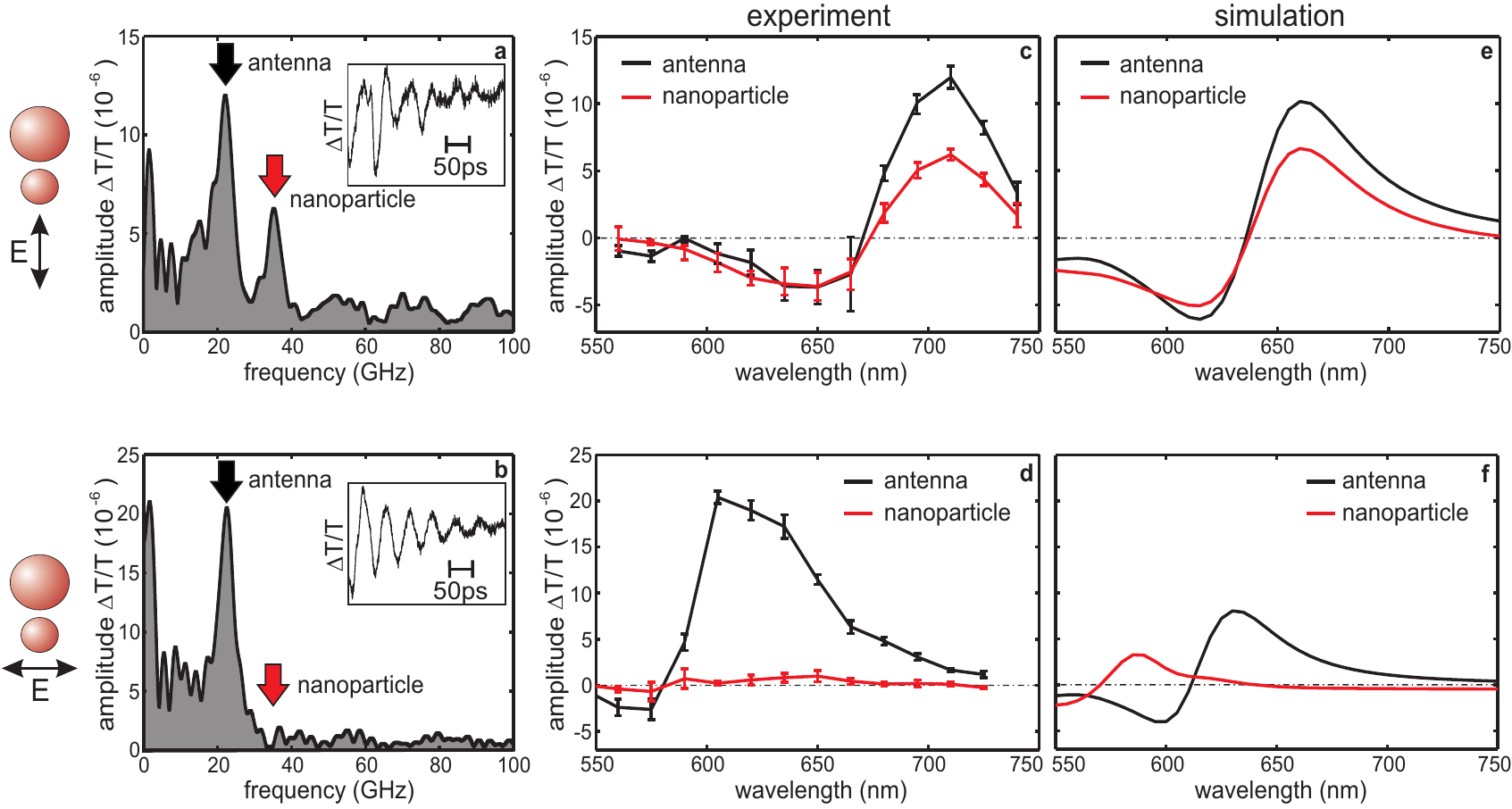}
\caption{\textbf{Experimental results compared to simulations:}  The
polarization of the probe light determines the coupling strength between antenna
and nanoparticle. The upper row of plots (\textbf{a,c,e}) shows the strong
coupling coupling case, lower row (\textbf{b,d,f}) the weak coupling control experiment. The insets
in \textbf{a,b} give an example for transient transmission traces, monitoring
the mechanical oscillation of the nanoparticles at their maximum signals. The
corresponding mechanical 
spectra (\textbf{a,b}) show always the antenna mode (22~GHz), but only in the
strong coupling case (\textbf{a}) also the nanoparticle mode at 36~GHz.
\textbf{c,d}, Oscillation amplitude of both modes as function of probe
wavelength as mean of six consecutive measurements. The error bars give the standard deviation. The line is a guide to the eye. \textbf{e,f}, Numerical simulations the oscillation amplitude
spectrum.} 
\label{fig:measured}
\end{figure*}

In the following we verify these theoretical predictions by ultrafast pump-probe
spectroscopy (for details see supplementary material) \cite{vandijk05dec31,hartland06annur}. A laser pulse is focused onto a single antenna-nanoobject pair. The pulse heats up the electron gas and subsequently the lattice. On a
picosecond timescale this launches mechanical breathing oscillations of the
particles. The periodic variation of the  particle volume causes a transient
time-dependent extinction.  After a variable time delay, a probe pulse 
measures the pump-induced relative transmission variation $\Delta T / T$
in the $10^{-6}$ range. By varying the time delay between pump- and  probe-pulse
we acquire mechanical oscillation traces.

The pump pulse excites the nanoparticle as well as the antenna. Consequently
both start to oscillate. The inset of Fig.{~\ref{fig:measured}a} shows an
example for the measured oscillation trace at 710~nm probe wavelength. Its
Fourier transform (Fig.{~\ref{fig:measured}a}) shows two distinct peaks at
22~GHz and 36~GHz. 
We assign the lower frequency mode to the antenna and the higher mode to the nanoparticle based on
measurements of individual single discs of different size.
The Fourier decomposition thus
allows us  to distinguish between the mechanical signals from antenna and
nanoparticle. In the following we plot the amplitudes of these two Fourier
components as a function of probe wavelength.

In the upper row of Fig.{~\ref{fig:measured}} the probe pulse is polarized along
the symmetry axis of the structure (strong antenna-nanoparticle coupling), and in the lower row perpendicular to it (weak coupling). In the first case (Fig.{~\ref{fig:measured}c}) the
oscillation amplitude spectra of antenna and nanoparticle are similar in
spectral shape and position. Both follow the coupled antenna-nanodisc mode. These findings agree well with a dipole-dipole coupling picture
(Fig.{~\ref{fig:idea}d and e}), in which the mechanical oscillation of the
smaller nanodisc causes a high-frequency modulation of the coupled mode extinction
signal. Full numerical simulations (see Fig.~\ref{fig:measured}e, calculation
details in the methods section) are in good agreement.  All spectral features
and signal amplitudes  as well as the amplitude ratio between antenna and
nanoparticle signal are well reproduced.

In order to perform a control experiment, we turn to the weak coupling case for a
probe polarization perpendicular to the symmetry axis. We  see a
significant change of the detected mechanical oscillation signal as plotted in
the inset of Fig.{~\ref{fig:measured}b} for a probe wavelength of 605~nm. The corresponding Fourier amplitudes 
show  just one peak, located at the mechanical
eigenmode frequency of the nanoantenna at  22~GHz. The  36 GHz signal of the
small nanoparticle is not recognizable, although the pump excitation process is
identical for both measurements. \textit{This behavior proves the concept of our
resonant antenna enhancement based on  polarization dependent plasmonic coupling.}
Plotting the mechanical oscillation signal of antenna and nanoparticle versus
the probe wavelength (Fig.{~\ref{fig:measured}d}), a clear change has
taken place when compared with the strong coupling polarization (Fig.{~\ref{fig:measured}c}). The oscillation
amplitude spectrum of the antenna is still clearly observable and located around
the uncoupled antenna resonance. The oscillation signal of the small
nanoparticle  vanishes almost completely.  The numerical model (Fig.{~\ref{fig:measured}f}) catches the essence: we observe a nearly undisturbed
signal from a single antenna and a very weak nanoparticle signal. 
\textit{Figures \ref{fig:measured}c and \ref{fig:measured}d demonstrate clearly
that we can turn the antenna enhancement on and off by switching the probe
polarization appropriately.}

\begin{figure}[t]
\centering
\includegraphics{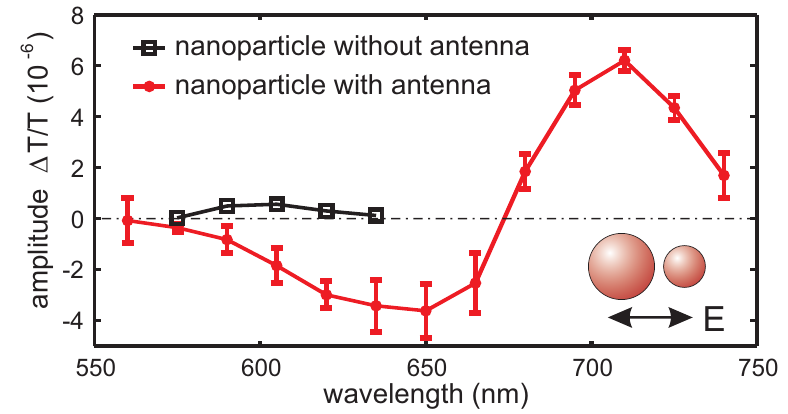}
\caption{\textbf{Enhancement of the oscillation signal by the nanoantenna:} The
weak signal measured on  a single  disc with 40~nm diameter (squares), compared
with the  signal amplified by the antenna structure (dots). As predicted we find
a red shift and an enhancement of the signal amplitude. The error bars give the
standard deviation of 6 consecutive measurements. The lines are guides to the
eye.}
\label{fig:singlevsantenna}
\end{figure}

Efficiency and spectral response are two fundamental features of an antenna. In order to
describe these features we have to compare the response of an antenna-enhanced
nanoparticle with that of a  single nanoparticle of the same size and shape. Due
to slight variations in size, crystallinity, and damping, the oscillation
amplitudes vary between nominally identical nanoparticles. Nanodiscs of 40~nm
diameter as used in combination with the antenna are at the limit of our current
detection capabilities. We increased the pump power and measured several discs
of that size (see supplementary material).  For the
comparison we select a disc in the center of the Gaussian amplitude distribution. The  oscillation amplitude
spectrum is plotted as squares in Fig.{~\ref{fig:singlevsantenna}}. 
It is obvious that the strongest signal is located around the
highest gradient of the plasmon resonance where the biggest extinction change is
expected. This reference signal can be compared to the antenna-enhanced signal
(dots in Fig.{~\ref{fig:singlevsantenna}}) which was already shown in
Fig.{~\ref{fig:measured}c. The feature is red-shifted and located around the
plasmon resonance of the coupled antenna. The maximum amplitude is about a
factor of ten larger than the signal from the nanoparticle without an antenna.
This mechanical oscillation signal is the strongest oscillation we have ever
measured of such a small single nanoparticle. As the antenna concentrates also
the pump field on the nanoparticle, stronger mechanical oscillations are
launched when the antenna is present. By numerical simulations we estimate an
increase by a factor of about 3, so that in the
present case pump and probe enhancement contribute equally to the total signal
enhancement.

We envision that our antenna-enhancement technique opens up new possibilities
for linear and nonlinear spectroscopy on the nanoscale. Not only plasmonic nanoparticles, but also  optically less active nanoobjects such  as dielectric particles and biomolecules will detune
the optical antenna. Their static linear as well as their ultrafast nonlinear response becomes thus accessible in the optical far-field.
We expect our method  not only to simplify the
research on single molecules or quantum dots, but also to open the way for modern
antenna-enhanced sensing and ultrafast telecommunication applications.

\vspace*{2\baselineskip}

\textbf{\hspace*{-1em}\large{Methods}} 
\paragraph*{\textbf{\hspace*{-1em}Experimental setup and data analysis}}
The output of a Ti:sapphire laser was used as pump pulses (wavelength 800~nm,
duration 200~fs, repetition rate 76~MHz). Probe pulses were generated by feeding
part of the pump pulses into a white-light fiber and selecting a 5~nm spectral
interval. This results in pulses of 2~ps duration tunable between 530 and
750~nm. After setting the polarization state, both pump and probe pulses are coupled
into a home-built sample-scanning confocal microscope and focused with a NA=0.95
objective on the sample surface. The transmitted light is collected using a
NA=1.3 oil immersion objective and detected by a balanced photodiode after
filtering out the pump light. Differential transmission traces were measured as a 
function of pump-probe delay.

To extract the oscillatory part, an exponential decay was fitted and subtracted,
removing the thermal background signal. The complex-valued Fourier transform was
calculated for each trace. The mechanical mode spectra
(Fig.{~\ref{fig:measured}a,b)} show the absolute Fourier values. The mechanical
oscillation signal of a given mode has a wavelength-independent phase relative
to the pump pulse, as all probe wavelengths monitor the same mechanical
oscillation \cite{perner00jul24,hodak99nov8}, causing a wavelength-independent
phase of a mode's complex-valued Fourier component  (for details see supplementary material). This is used to check the
quality of the data and to retrieve the sign in the amplitude spectrum
Fig.{~\ref{fig:measured}c,d} and Fig.{~\ref{fig:singlevsantenna}}.

\paragraph*{\textbf{\hspace*{-1em}Numerical simulations}}
For the numerical simulations we used a Null-Field Method \cite{tmatrixbook} and
a  Finite-Element Method \cite{comsol} which give nearly equal results. We
approximated our structure by gold nanodiscs in an homogeneous medium.  For a
single disc, an effective index of refraction of n=1.42 yields a good agreement
between measured and calculated dark field scattering spectra. However, the mode
splitting of a coupled disc pair is underestimated in this model. The step in
index of refraction at the interface seems to be more important in near-field
coupling. This explains the differences in spectral position between experiment
and simulation in Fig.{~\ref{fig:measured}}. For the differential transmission
data we took the difference between an unperturbed sample and a sample in which
both the nanoparticle's dielectric function as well as its size was changed. For
the size change of the particle we use an isotropic thermal expansion model, and
a  temperature variation proportional to the Joule heating of the pump pulse in
the particle. The change of the dielectric properties is modeled in our
wavelength range by a Drude model. The size-induced variation of the electron
density results in a variation of the plasma frequency \cite{ergin09jun}. The
d-band absorption is extracted from Johnson and Christy data
\cite{johnson-christy72} and kept constant. This renders  our simulations 
below a probe wavelength of about 600~nm unreliable.

%%%%%%%%%%%%%%%%%%%%%%%%%%%%%%%%%%%%%%%%%%

%%%%%%%%%%%%%%%%%%%%%%%%%%%%%%%%%%%%%%%%%%%%%

\vspace*{1\baselineskip}

\textbf{\hspace*{-1em}\large{Acknowledgments}} 
\newline
We gratefully acknowledge financial support from the 
Baden-W\"urttemberg Stiftung, the Deutsche Forschungsgemeinschaft and BMBF.

\vspace*{1\baselineskip}

\textbf{\hspace*{-1em}\large{Author contributions}} 
\newline
M.L. designed the experiment. T.S. and K.K. conducted the experiments and simulations. D.M. worked on the project at an early phase. M.H. prepared the sample. T.S., H.G. and M.L. wrote the paper.


\begin{thebibliography}{10}
\bibitem{footnoteresonant}
\bibinfo{title}{With resonant we mean that the antennas possess a plasmonic resonance at the wavelength of operation in contrast to purely scattering near-field tips.}
\end{thebibliography}
\end{document}